%% file: surf_charge.tex
\newif\iflanl
\lanlfalse
\lanltrue

\iflanl
\documentclass[superscriptaddress,twocolumn,footinbib,bibnotes,oneside,prb,showkeys,showpacs,amsmath,amsfonts]{revtex4}
\else
\documentclass[superscriptaddress, preprint,footinbib,bibnotes,oneside,prb,noshowkeys,noshowpacs,amsmath,amsfonts]{revtex4}
\fi

\newif\ifpdf 
\ifx\pdfoutput\undefined
\pdffalse 
\else
\pdfoutput=1 
\pdftrue
\fi

\ifpdf 
\usepackage[pdftex]{graphicx}
\usepackage[pdftex]{hyperref}
\usepackage{thumbpdf}
\else 
\usepackage{graphicx}
\usepackage[dvips]{hyperref}  
\fi

\hypersetup{pdfsubject={Preprint; Journal submission},   
  pdftitle  = {Potential and field singularity at a surface point charge},
  pdfauthor = {Alexander Silbergleit, Ilya Mandel, and Ilya Nemenman},
  pdfkeywords = {point charge, surface charge, fluxon, Laplace equation,
    Neumann boundary value problem, singularity},
  bookmarks ={true}, 
}

\include{aliases}

\begin{document}

\iflanl
\else
\def\bibsection{\newpage\section*{References}}
\renewcommand{\contentsline}[4]{\hbox to\textwidth{Fig.~#2\hfill page #3}}
\def\lofname{Figure captions}
\fi

\title{Potential and field singularity at a surface point charge
\pdfbookmark[1]{Title page}{tit}
}

\author{Alexander Silbergleit} \email{gleit@relgyro.stanford.edu}
\affiliation{Gravity Probe B, W.\ W.\ Hansen Experimental Physics
  Laboratory, \\Stanford University, Stanford, CA 94305-4085, USA}

\author{Ilya Mandel} \email{ilya@caltech.edu} \affiliation{Physics
  Department, Mail Code 103-33, California Institute of Technology,
  Pasadena, CA 91125, USA}

\author{Ilya Nemenman} \email{nemenman@kitp.ucsb.edu}
\affiliation{Kavli Institute for Theoretical Physics, University of
  California, Santa Barbara, CA 93106, USA}

\begin{abstract}
  The behavior of the magnetic potential near a point charge (fluxon)
  located at a curved regular boundary surface is shown to be
  essentially different from that of a volume point charge. In
  addition to the usual inverse distance singularity, two singular
  terms are generally present. The first of them, a logarithmic one,
  is axially symmetric with respect to the boundary normal at the
  charge location, and proportional to the sum of the two principal
  curvatures of the boundary surface at this point, that is, to the
  local mean curvature. The second term is asymmetric and proportional
  to the difference of the two principal curvatures in question; it is
  also bounded at the charge location.  Both terms vanish, apparently,
  if the charge is at a planar point of the boundary, and only in this
  case. The field in the charge vicinity behaves accordingly,
  featuring generally two singular terms proportional to the inverse
  distance, in addition to the main inverse distance squared
  singularity. This result is significant, in particular, for studying
  the interaction of magnetic vortices in type II superconductors.
\end{abstract}



\preprint{NSF-KITP-03-37}

\date{\today} \maketitle

\pdfbookmark[1]{Article body}{ab}
Magnetic vortex lines are formed in type II superconductors
\cite{tink}.  When crossing the superconductor boundary, they create
strongly localized surface sources of magnetic field (fluxons), which
may play an important role in various physical situations. For
instance, two space tests of Einstein's General Relativity, Gravity
Probe B\cite{everitt-88,bard} and STEP (Space Test of the Equivalence
Principle)\cite{mester-01}, are based on low temperature technology
with type II superconductors, and their setup is significantly
affected by fluxons.

The size of a surface magnetic spot is about the microscopic London
length~\cite{tink}, i.~e., it is typically much smaller than
characteristic macroscopic sizes involved. Thus the {\em point charge}
approximation appears naturally and proves to be sufficient for many
applications. Within this approximation, the magnetic potential,
$\psi=\psi({\bf R})$, satisfies the Neumann boundary value problem
\begin{eqnarray}
  \Delta \psi &=&0,\quad {\bf R}\in D\;,
  \label{eq:laplace}\\
  \frac{\partial\psi}{\partial n}\biggl|_S &=&\,\sum_{j=1}^{N}\nu_j\,\Phi_0\;
  \delta_S ({\bf R}-{\bf R}_j),\quad
  {\bf R},\,{\bf R}_j\in S\,.
  \label{eq:boundary}
\end{eqnarray}
Here the domain $D$ is the empty space, surface $S$ is the
superconductor boundary, $\Phi_0 = h/2e$ is the magnetic flux quantum
in SI units, and the magnetic field is ${\bf B}=-\nabla \psi$.
Moreover, $\delta_S ({\bf R}-{\bf R}_j)$ denotes the surface
delta-function at the position, ${\bf R}_j$, of a vortex, and $\nu_j$
is either plus or minus one, depending on whether the field line
enters the domain $D$ ($\nu_j=+1$), or exits it ($\nu_j=-1$). We
assume that the boundary $S$ is smooth enough (at least $C^3$) near
every charge.  Outside the charge vicinities it may have any
singularities compatible with the finite local energy condition,
meaning $(\nabla \psi)^2$ is locally integrable.

If $D$ is bounded, then each vortex line starts and ends at the
boundary, the number of charges is even, and the total charge
vanishes, $\sum_{j=1}^{N}\nu_j\,\Phi_0=0$, which condition is the
solvability criterion of the problem,
Eqs.~(\ref{eq:laplace},~\ref{eq:boundary}). If the domain $D$ is
infinite, some field lines may end at infinity, and this condition may
not hold; in any case, we do not use it in the following analysis,
which is entirely local.

An immediate question regarding the above boundary value problem is
how does its solution behave near a surface charge? For a curved
boundary, an answer based on the similarity with the volume point
charge turns out incorrect.  This is seen from the simplest example, a
spherical domain.  A closed--form exact solution to
Eqs.~(\ref{eq:laplace},~\ref{eq:boundary}) in the exterior of a sphere
was obtained in Ref.~\onlinecite{ns-99}. It shows that a new
logarithmic singular term, inversely proportional to the radius of the
sphere, is added to the main inverse distance singularity in the
expansion of the potential near the charge. So, what happens with the
singularity for a generally curved smooth surface?

Our search for the answer to this natural and, in fact, classical
question covered books and papers in both mathematical physics and in
the field of vortices in superconductors, as well as communications
with colleagues in both fields. We also talked with high energy
theorists expecting to find perhaps some relevant results in view of
the discussions of the magnetic monopole. However, no ready answer was
found, which might be not so surprising. Indeed, the Neumann boundary
value problem with surface charges is not relevant to the design of
electrostatic systems. On the other hand, its magnetostatic
implementation became available only with the widespread technical use
of superconductors in the recent years. Last but not least, the answer
proves to be not that simple.

In this letter we fill the gap by deriving a complete singular part of
the expansion of the solution to
Eqs.~(\ref{eq:laplace},~\ref{eq:boundary}) near a charge at an
arbitrary curved smooth boundary. As compared to the case of a sphere,
one more singular term, proportional to the difference of the two
principal curvatures, appears in the general case.

\begin{figure}
  \centerline{\includegraphics[width=3.2in]{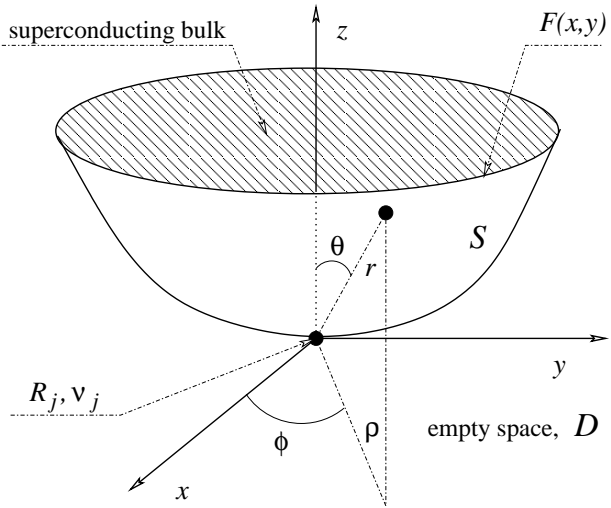}}
  \caption{\label{fig:near}Definition of coordinate systems near a charge.}
\end{figure}

We are interested in the behavior of the potential near a single
surface charge at some ${\bf R}_j$. For brevity, we thus drop the
charge index in the following calculation.  We put the origin of a
Cartesian coordinate system at ${\bf R}_j$, so that ${\bf r}\equiv{\bf
  R}-{\bf R}_j$. We point the $z$ axis along the outward normal to the
surface $S$ (that is, into the superconducting bulk), choosing the $x$
and $y$ axes in the tangent plane, so that the unit vectors $\{\hat x,
\hat y, \hat z\}$ form a right orthogonal triplet. Along with
Cartesian $\{x,\,y,\,z\}$, we will use the corresponding spherical,
$\{r,\,\theta,\,\phi\}$, and cylindrical, $\{\rho,\,\phi,\,z\}$,
coordinate systems (see~Fig.~\ref{fig:near}).

The shape of the smooth boundary surface in the vicinity of the charge
can be described by the equation $z=F(x,y)$. The Taylor expansion of
the function $F(x,y)$ around $x=y=0$ apparently has no terms linear in
$x$ or $y$, since $z$ is oriented along the normal.  Moreover, by an
appropriate rotation of the coordinate axes $\hat x, \hat y$ in the
tangent plane, we can ensure that the second cross-derivative of $F$
vanishes at the origin, hence the expansion acquires the form
\begin{equation}\label{3}
{z}= F(x,y)
=\frac{k^{(x)}}{2}x^2 + \frac{k^{(y)}}{2}y^2
+O(\rho ^3)\equiv f(x,y)+O(\rho ^3)\,,
\end{equation}
where
\begin{equation}\label{4}
k^{(x)}=\frac{\partial^2 F}{\partial x^2} \biggl|_{x= y=0}, \qquad
k^{(y)}=\frac{\partial^2 F}{\partial y^2} \biggl|_{x= y=0}
\end{equation}
are the two principal curvatures of the boundary surface at the charge
location.

Since, near ${\bf r}=0$,
\begin{eqnarray*}
  \delta_S ({\bf r})&=&\delta (x)\delta ({y})/J,\qquad\\ 
  {\partial}/{\partial n}&=&\hat n\cdot\nabla=
  (1/J) 
  \left(
    {\partial}/{\partial z}-F_x\,{\partial}/
    {\partial x}-F_y\,{\partial}/{\partial y}
  \right)\; ;\qquad\\
  J&\equiv&\sqrt{1+F_x^2+ F_y^2}\; ,
\end{eqnarray*}
the boundary condition, Eq.~(\ref{eq:boundary}), in the vicinity of
the charge can be written in terms of variables $x, y, z$ as
\footnote{Although the Neumann boundary condition should not, in
  general, allow for a solution for a single vortex, we avoid this
  complication by extending below the domain $D$ away from the
  fluxon.}:
\iflanl
\begin{multline}\label{5}
\frac{\partial\psi}{\partial z}\biggl|_{z=F(x,y)} =
\nu \Phi_0 \delta(x) \delta(y)
\\+
\left(F_x\,\frac{\partial\psi}
 {\partial x}+F_y\,\frac{\partial\psi}
 {\partial y}\right)\biggl|_{z=F(x,y)}\; .
\end{multline}
\else
\begin{equation}\label{5}
\frac{\partial\psi}{\partial z}\biggl|_{z=F(x,y)} =
\nu \Phi_0 \delta(x) \delta(y)+
\left(F_x\,\frac{\partial\psi}
 {\partial x}+F_y\,\frac{\partial\psi}
 {\partial y}\right)\biggl|_{z=F(x,y)}\; .
\end{equation}
\fi
The partial derivatives of the function $F(x,y)$ near the origin are
given, to the order we are interested in, by
\begin{equation}\label{Fxy}
F_x={k^{(x)}}x + O(\rho ^2),\qquad F_y={k^{(y)}}y + O(\rho ^2)\; .
\end{equation}

Once again, we only care about the vicinity of the charge where
$z=F(x,y)$ is small, so we can use perturbation of the boundary to
move the boundary condition, Eq.~(\ref{5}), to the plane $z=0$. This
is done by means of the following Taylor expansion of an arbitrary
function $w=w(x,y,z)$:
\[
w\biggl|_{z=F(x,y)} = w\biggl|_{z=0}+ F\,\frac{\partial w}{\partial
  z}\biggl|_{z=0} + \frac{F^2}{2}\,\frac{\partial^2 w}{\partial
  z^2}\biggl|_{z=0} + \ldots\; .
\]
Applying this to the derivatives of $\psi$ in Eq.~(\ref{5}) we write
it, to the proper order, in the form:
\iflanl
\begin{multline}\label{6}
\frac{\partial\psi}{\partial z}\biggl|_{z=0} = 
\nu \Phi_0 \delta(x) \delta(y)
\\
+\left(F_x\,\frac{\partial\psi}{\partial x}+
F_y\,\frac{\partial\psi}{\partial y}-F\,\frac{\partial^2 \psi}{\partial z^2}
\right)
\biggl|_{z=0}+\dots\; .
\end{multline}
\else
\begin{equation}\label{6}
\frac{\partial\psi}{\partial z}\biggl|_{z=0} = 
\nu \Phi_0 \delta(x) \delta(y)+\left(
F_x\,\frac{\partial\psi}{\partial x}+
F_y\,\frac{\partial\psi}{\partial y}-F\,\frac{\partial^2 \psi}{\partial z^2}
\right)
\biggl|_{z=0}+\dots\; .
\end{equation}
\fi

The final step of this derivation is to expand $\psi$ in a series of
successively smaller (that is, less singular at the origin) functions
$\psi^{(i)}$,
\begin{equation}\label{eq:psiexpansion}
\psi=\psi^{(0)}+\psi^{(1)}+\psi^{(2)}+...\; .
\end{equation}
Introducing this expansion in the boundary condition Eq.~(\ref{6}) and
then matching the terms of the same order, we end up with the
following sequence of boundary conditions for $\psi^{(i)},\,\,
i=0,1,2,\dots,$ at $z=0$: 
\iflanl
\begin{eqnarray}
\label{eq:psi0}
\frac{\partial\psi^{(0)}}{\partial z}
\biggl|_{z=0}&=&\nu \Phi_0 \delta(x)\delta(y)\; ;
\\
\label{eq:psi1}
\frac{\partial\psi^{(1)}}{\partial z}\biggl|_{z=0}&=&
\left[
k^{(x)}x\,\frac{\partial\psi^{(0)}}
{\partial x} +k^{(y)}y\,\frac{\partial\psi^{(0)}}{\partial y}
\right.\nonumber\\&&\left.-
f(x,y)\,\frac{\partial^2 \psi^{(0)}}{\partial z^2}
\right]
\biggl|_{z=0} \; ,
\end{eqnarray}
\else
\begin{eqnarray}
\label{eq:psi0}
\frac{\partial\psi^{(0)}}{\partial z}
\biggl|_{z=0}&=&\nu \Phi_0 \delta(x)\delta(y)\; ;
\\
\label{eq:psi1}
\frac{\partial\psi^{(1)}}{\partial z}\biggl|_{z=0}&=&
\left[
k^{(x)}x\,\frac{\partial\psi^{(0)}}
{\partial x}+k^{(y)}y\,\frac{\partial\psi^{(0)}}{\partial y}-
f(x,y)\,\frac{\partial^2 \psi^{(0)}}{\partial z^2}
\right]
\biggl|_{z=0} \; ,
\end{eqnarray}
\fi
and so on. Here we have dropped higher order terms in the right-hand
sides by replacing $F$ and $F_x,\;F_y$ with their main term
expressions from Eqs.~(\ref{3}) and (\ref{Fxy}), respectively. Of
course, all functions $\psi^{(i)}$ are subject to the Laplace
equation, Eq.~(\ref{eq:laplace}).

Thus, {\it locally} we have successfully replaced the boundary value
problem of Eqs.~(\ref{eq:laplace},~\ref{eq:boundary}) in the domain
$D$ by a sequence of problems for functions $\psi^{(i)},\,\,
i=0,1,2,\dots$, harmonic in the half--space $z<0$ and satisfying the
above boundary conditions, Eqs.~(\ref{eq:psi0}),~(\ref{eq:psi1}), etc.
We now need to solve these problems for the half--space one by one,
until the normal derivative of the solution becomes finite at the
boundary.

The zero order solution $\psi^{(0)}$ obeying the boundary condition of
Eq.~(\ref{eq:psi0}) is, of course,
\begin{equation}\label{eq:zeroorder}
\psi^{(0)} =  \frac{\nu\Phi_0}{2\pi} \frac{1}{r}\; .
\end{equation}
It allows one to immediately calculate the r.\ h.\ s.\ of
Eq~(\ref{eq:psi1}). Indeed,
\iflanl
\begin{eqnarray*}
f\frac{\partial^2\psi^{(0)}}{\partial z^2}\biggl|_{z=0}&=&
-\frac{\nu\Phi_0f}{2\pi}\left(\frac{1}{r^3}-
  \frac{3z^2}{r^5}\right)\biggl|_{z=0}\\&=&
-\frac{\nu\Phi_0}{2\pi}\frac{k^{(x)} x^2 + k^{(y)} y^2}{2\rho^3} \; ,
\end{eqnarray*}
\else
\[
f\frac{\partial^2\psi^{(0)}}{\partial z^2}\biggl|_{z=0}=
-\frac{\nu\Phi_0f}{2\pi}\left(\frac{1}{r^3}-
  \frac{3z^2}{r^5}\right)\biggl|_{z=0}=
-\frac{\nu\Phi_0}{2\pi}\frac{k^{(x)} x^2 + k^{(y)} y^2}{2\rho^3} \; ,
\]
\fi
where the second term in the middle expression turns to zero at $z=0$,
contributing no $\delta$--like singularities, due to the presence of
the factor $f=O(\rho^2)$.  Taking also into account that ${\partial
  r^{-1}}/{\partial x}=-x/r^{3}$, ${\partial r^{-1}}/{\partial
  y}=-y/r^{3}$, we find the boundary condition for $\psi^{(1)}$ in its
final explicit form:
\iflanl
\begin{eqnarray}
\frac{\partial\psi^{(1)}}{\partial z}\biggl|_{z=0}&=&
-\frac{\nu\Phi_0}{2\pi}\frac{k^{(x)} x^2 + k^{(y)} y^2}{2\rho^3}
\nonumber \\&=& 
-\frac{\nu\Phi_0}{8\pi}
\biggl[\frac{k^{(x)}+k^{(y)}}{\rho}\nonumber\\ &&+\frac{k^{(x)}-k^{(y)}}
{\rho}\cos{2\phi}\biggr]\,.
\end{eqnarray}
\else
\begin{equation}
\frac{\partial\psi^{(1)}}{\partial z}\biggl|_{z=0}=
-\frac{\nu\Phi_0}{2\pi}\frac{k^{(x)} x^2 + k^{(y)} y^2}{2\rho^3}
\\=
-\frac{\nu\Phi_0}{8\pi}
\biggl[\frac{k^{(x)}+k^{(y)}}{\rho}+\frac{k^{(x)}-k^{(y)}}
{\rho}\cos{2\phi}\biggr]\,.
\end{equation}
\fi
The two terms on the utmost right here have essentially different
singularities at the origin. For this reason, we treat them separately
by splitting the problem in two in the following way:
\begin{eqnarray}
\psi^{(1)}&=&\psi^{(1)}_s + \psi^{(1)}_r,\label{eq:psi_sr}
\\
\frac{\partial\psi^{(1)}_s}{\partial z}\biggl|_{z=0}&=&
-\frac{\nu\Phi_0}{8\pi}\frac{k^{(x)}+k^{(y)}}{\rho}
\label{eq:bound_s}
\\
\frac{\partial\psi^{(1)}_r}{\partial z}\biggl|_{z=0}&=&
-\frac{\nu\Phi_0}{8\pi}\frac{k^{(x)}-k^{(y)}}{\rho}\cos{2\phi}\; .
\label{eq:bound_r}
\end{eqnarray}

The Neumann problem for $\psi^{(1)}_s$ in the half--space does not
have solutions bounded at infinity, as one would expect in our
investigation (we are actually looking for terms {\it growing} away
from the charge, because a weaker singularity next to the inverse
distance is most probably some logarithm tending to infinity at both
the charge and the infinite distance from it). For this reason, no
solution can be found by means of standard techniques. However, a
harmonic and regular in the half--space $z<0$ function
\iflanl
\begin{eqnarray}\label{eq:psis1}
\psi^{(1)}_{s}&=&  K_+\ln [(r-z)/d] \nonumber \\
&=&
  K_+\left[\ln (r/d) + \ln(1-\cos\theta)\right],\qquad \\
K_{\pm}&\equiv&\nu\Phi_0\left[k^{(x)}\pm k^{(y)}\right]/8\pi\; ,
\label{eq:Kpm}
\end{eqnarray}
\else
\begin{eqnarray}\label{eq:psis1}
\psi^{(1)}_{s}&=&  K_+\ln [(r-z)/d] =
  K_+\left[\ln (r/d) + \ln(1-\cos\theta)\right],\qquad \\
K_{\pm}&\equiv&\nu\Phi_0\left[k^{(x)}\pm k^{(y)}\right]/8\pi\; ,
\label{eq:Kpm}
\end{eqnarray}
\fi
where $d>0$ is an arbitrary constant of the dimension of length,
provides the needed solution. Indeed, it satisfies the boundary
condition, Eq.~(\ref{eq:bound_s}), in view of
\iflanl
\[
\frac{\partial\ln(r-z)}{\partial z} = \frac{{z}/{r}-1}{r-z}=
-{1}/{r}\to -{1}/{\rho},\qquad z\to -0\;.
\]
\else
\[
{\partial\ln(r-z)}/{\partial z} = ({r-z})^{-1}\left({z}/{r}-1\right)=
-{1}/{r}\to -{1}/{\rho},\qquad z\to -0\;.
\]
\fi
The solution given by Eq.~(\ref{eq:psis1}) is unique in the class of
functions with the logarithmic growth at infinity, namely, those with
the asymptotics
\iflanl
\begin{eqnarray*}
\psi^{(1)}_s&=& K_+\ln (r/d) + K_+\ln (1-\cos\theta)+o(1),\quad\\
\frac{\partial \psi^{(1)}_s}{\partial r}&=& K_+/r+O(1/r^2),\qquad
r\to\infty\; .
\end{eqnarray*}
\else
\[
\psi^{(1)}_s= K_+\ln (r/d) + K_+\ln (1-\cos\theta)+o(1),\quad
\frac{\partial \psi^{(1)}_s}{\partial r}= K_+/r+O(1/r^2),\qquad
r\to\infty\; .
\]
\fi

Contrary to the previous one, the Neumann problem for $\psi^{(1)}_r$,
\begin{equation}\label{14}
\Delta\psi^{(1)}_r=0,\quad z<0,\qquad
\frac{\partial\psi^{(1)}_r}{\partial z}\biggl|_{z=0}=
-\frac{K_{-}}{\rho}\cos{2\phi}\; ,
\end{equation}
has a unique, up to a constant, solution bounded at infinity [namely,
a solution that obeys somewhat unusual conditions
$\psi^{(1)}_r=O(1),\quad {\partial \psi^{(1)}_r}/{\partial
  r}=o(1/r^2),\quad r\to\infty$]. The solution is obtained by the
standard separation of variables in cylindrical coordinates using the
Hankel transform, and it reads:
\begin{eqnarray}\label{eq:hankel}
  \psi^{(1)}_r &=& -K_{-}\cos{2\phi}\,\int_0^\infty J_2(\lambda \rho)
  \exp(-\lambda |z|)\, \frac{d\lambda}{\lambda} \nonumber\\&=&
  -\frac{K_{-}\cos{2\phi}}{2}\,\left(\frac{\rho}{r-z}\right)^2 
\iflanl\nonumber\\&=&
\else=
\fi
  -\frac{K_{-}}{2}\,\frac{x^2-y^2}{(r-z)^2} \; .
\end{eqnarray}
The value of the integral is found in Ref.~\onlinecite{int-tran-1},
4.14.(5), and the constant $K_{-}$ is defined in Eq.~(\ref{eq:Kpm}).
Interestingly, this solution in spherical coordinates does not depend
on the radius, being a function of the angles only [singular on the
positive semi--axis $z>0$, same as $\psi^{(1)}_s$ in
Eq.~(\ref{eq:psis1})]:
\[
\psi^{(1)}_r =
-\frac{K_{-}}{2}\,\frac{\sin^2\theta\,\cos{2\phi}}
{(1-\cos\theta)^2},\qquad\qquad
\frac{\partial \psi^{(1)}_r}{\partial r} = 0\; .
\]

It is now straightforward to see that the Neumann boundary data for
all higher order corrections to the potential, starting with
$\psi^{(2)}$, are finite at the origin (and dropping fast enough at
infinity); accordingly, the solutions of the corresponding problems
bounded at infinity are unique up to an additive constant. It also
means that all the terms in the expansion,
Eq.~(\ref{eq:psiexpansion}), of the potential, whose normal derivative
are singular at the location of a surface charge, are given by the
solutions already found. Hence, combining the expressions from
Eqs.~(\ref{eq:zeroorder}),~(\ref{eq:psis1}), and~(\ref{eq:hankel}), we
find the desired formula for the magnetostatic potential near a
surface charge ($r\to0$):
\iflanl
\begin{eqnarray}
  \psi&=&\psi^{(0)}+\psi^{(1)}_s+\psi^{(1)}_r+...\nonumber\\
  &=&\frac{\nu\Phi_0}{2\pi } \biggl[ \frac{1}{r} +
  \frac{k^{(x)}+k^{(y)}}{4}\ln\frac{r-z}{d} \nonumber\\
  &&-
  \frac{k^{(x)}-k^{(y)}}{8}\frac{x^2-y^2}{(r-z)^2} \biggr]+
  \parbox{.8in}{(nonsingular terms)}\; .
\label{eq:psi_solved}
\end{eqnarray}
\else
\begin{eqnarray}
  \psi&=&\psi^{(0)}+\psi^{(1)}_s+\psi^{(1)}_r+...\nonumber\\
  &=&\frac{\nu\Phi_0}{2\pi } \biggl[ \frac{1}{r} +
  \frac{k^{(x)}+k^{(y)}}{4}\ln\frac{r-z}{d} -
  \frac{k^{(x)}-k^{(y)}}{8}\frac{x^2-y^2}{(r-z)^2} \biggr]+
  \mbox{(nonsingular terms)}\; .
\label{eq:psi_solved}
\end{eqnarray}
\fi
It is easy to rewrite this in our general notations from
Eqs.~(\ref{eq:laplace},~\ref{eq:boundary}) by replacing $|{\bf r}|$
with $|{\bf R}-{\bf R}_j|$, $x$ with $X-X_j$, etc. Instead, we give
the expression of the singular part of the magnetic field near the
charge. It can be written in the form:
\iflanl
\begin{eqnarray}
\label{field}
{\bf B}&=&-\nabla\psi\nonumber\\
&=& \frac{\nu\Phi_0}{2\pi } \left[ \frac{\hat r}{r^2}
  -\frac{k^{(x)}+k^{(y)}}{4r} \left( \hat
    r+\frac{\sin\theta}{1-\cos\theta}\hat\theta
  \right)\right.\nonumber\\
&&-\left.\frac{k^{(x)}-k^{(y)}}{4r}\frac{\sin\theta}{(1-\cos\theta)^2}
  \left( \cos2\phi\,\hat\theta+\sin2\phi\,\hat\phi \right) \right] \nonumber\\&&+
\parbox{1.5in}{(nonsingular terms)} \; .
\end{eqnarray}
\else
\begin{eqnarray}
\label{field}
{\bf B}&=&-\nabla\psi= \frac{\nu\Phi_0}{2\pi } \left[ \frac{\hat r}{r^2}
  -\frac{k^{(x)}+k^{(y)}}{4r} \left( \hat
    r+\frac{\sin\theta}{1-\cos\theta}\hat\theta
  \right)\right.\nonumber\\
&&-\left.\frac{k^{(x)}-k^{(y)}}{4r}\frac{\sin\theta}{(1-\cos\theta)^2}
  \left( \cos2\phi\,\hat\theta+\sin2\phi\,\hat\phi \right) \right] +
\parbox{1.5in}{(nonsingular terms)} \; .
\end{eqnarray}
\fi

Here are a few concluding remarks regarding the obtained result.

First, notice that the leading order contribution to the potential,
Eq.~(\ref{eq:zeroorder}), is twice that of the point charge located in
a volume away from its boundaries. This is clearly explained by the
fact that the field lines and the flux from the surface charge emanate
only into the half--space, versus the full space for the volume
charge.

The two singular corrections to the usual inverse distance singularity
of the potential, Eq.~(\ref{eq:psi_solved}), are very different. The
first one is logarithmic, symmetric about the direction of the normal
to the boundary at the charge location, and proportional to the sum of
two principal surface curvatures there, i.~e., to the mean boundary
curvature. Thus, it vanishes if the charge sits at a symmetric saddle
point of the boundary. The second additional singularity is
asymmetric, proportional to the difference of the principal
curvatures, and vanishes thus when the latter are equal, i.~e., when
the charge is at a spherical point of the boundary. This second term
is bounded at the charge location [giving unbounded field components,
see Eq.~(\ref{field})], but is not uniquely defined there, with the
limiting values depending on the direction along which the limit is
taken. Note that both corrections vanish simultaneously if and only if
the charge is at the planar point of the boundary.

In a particular case when the domain $D$ is the exterior of a sphere
of the radius $a$, one has $k^{(x)}=k^{(y)}=1/a$. If there is just one
surface charge, $N=j=1$ and $\nu=1$ (so that the incoming vortex line
ends at infinity), the Eq.~(\ref{eq:psi_solved}) becomes
\iflanl
\begin{multline}
  \psi=  \frac{\Phi_0}{2\pi } \left[
\frac{1}{|{\bf R}-{\bf R}_1|} +
   \frac{1}{2a}\ln\frac{|{\bf R}-{\bf R}_1|-
     \hat{n}\cdot ({\bf R}-{\bf R}_1)}{d}\right]
  \\+ \parbox{2.0in}{(nonconstant nonsingular terms)}
\; ,
\end{multline}
\else
\begin{equation}
  \psi=  \frac{\Phi_0}{2\pi } \left[
\frac{1}{|{\bf R}-{\bf R}_1|} +
   \frac{1}{2a}\ln\frac{|{\bf R}-{\bf R}_1|-
     \hat{n}\cdot ({\bf R}-{\bf R}_1)}{d}\right]
  + \parbox{1.5in}{(nonconstant nonsingular terms)}
\; ,
\end{equation}
\fi
in complete agreement with the exact solution obtained in
Ref.~\onlinecite{ns-99} with $d=2a$.

Finally, the obtained singular expansion of the potential can be used
in the derivation of the force acting on a charge in a fashion similar
to the one developed in the case of volume point
charges~\cite{snm-03}, i.~e., by means of the geometrical
regularization of energy and, henceforth, the force, as the energy
gradient in the charge location. However, in a striking contrast with
the volume case, the force here is found to depend on the gradient of
the curvature at the charge location. Namely, due to the first
additional singular term in the potential, Eq.~(\ref{eq:psi_solved}),
there appears a tangential force on the charge which tries to move it
towards the point of the stationary mean curvature of the boundary,
and which diverges in the regularization limit. If confirmed, this
divergence would mean that either the approximation of the {\it point}
surface charges does not completely describe real microscopic, but
finite size fluxons, or, strangely enough, that the fluxons cannot
reside at arbitrary points of a curved boundary, or perhaps even
something else.

A detailed study of fluxon interactions will be carried out in a
separate publication. However, it is clear that it will necessarily
use the results of this paper, in view of the relation
\begin{equation}
\psi({\bf r},\mu) = \int_{S} dS(\bxi) \, \mu(\bxi) \,
\psi({\bf r} - \boldsymbol{\xi})\,,
\end{equation}
where $\psi({\bf r},\mu)$ is the potential created by the surface
charge density $\mu(\bxi), \xi\in S$, and $\psi({\bf r})$ is the
potential from Eq.~(\ref{eq:psi_solved}). For small, yet finite size
fluxons the divergent asymptotics derived above will have an explicit
short scale cutoff defined by the spatial extent of the density
(presumably, the London length).  However, the detailed analysis will
require a deeper insight in the real structure of magnetic vortex
lines near a boundary. Without such an analysis one cannot, in fact,
speculate about the strength and importance of these surface
interactions; we will thus limit ourselves to just a few short
comments.

First, one compares, naturally, the surface force coming from the
logarithmic term in the field potential to the strength of the random
pinning force that defines the fluxon's position \cite{blatter-94}.
The latter depends on the flux tube length and the former does not.
So, allowing for a physical regularization of the mathematically
divergent surface effects, one will in any case come up with some
characteristic length, $L$, below which the surface force will
dominate.  The description of the vortex line dynamics that does not
account for surface effects at distances from the surface smaller than
$L$ is necessarily incomplete.

Second, forces between two vortices in a superconducting bulk are
exponentially small if the vortex line separation is larger than the
London length (precisely the regime we are discussing). These forces
can be neglected. Thus, the surface effects we have found will be the
leading interaction terms.  Such effects are significant and translate
into an experimentally relevant magnetic ``friction'' between
superconducting bodies \cite{glen}.

\begin{acknowledgments}
  We thank Leonid Bakaleinikov, David Gross, Lev Kapitanski, Akakii
  Melikidze, Andrei Ruckenstein, Andrey Shytov, and Robert Wagoner for
  discussions and valuable references. A.~S.\ and partly I.~M.\ were
  supported by NASA grant NAS 8-39225 to Gravity Probe B. I.~N.\ was
  supported by NSF grant PHY99-07949 to Kavli Institute for
  Theoretical Physics.
\end{acknowledgments}

\iflanl
\else
\vfill\eject
\listoffigures
\fi

\end{document}

%% file: aliases.tex
\newcommand{\bde}{\begin{description}}

\newcommand{\ben}{\begin{enumerate}}
\newcommand{\beq}{\begin{eqnarray}}

\newcommand{\beqn}{\begin{eqnarray*}}

\newcommand{\bqu}{\begin{quote}}

\newcommand{\bvb}{\begin{verbatim}}

\newcommand{\bxi}{\boldsymbol{\xi}}

\newcommand{\ea}{\end{eqnarray}}

\newcommand{\ede}{\end{description}}
\newcommand{\een}{\end{enumerate}}
\newcommand{\eeq}{\end{eqnarray}}
\newcommand{\eeqn}{\end{eqnarray*}}

\newcommand{\equ}{\end{quote}}

\newcommand{\evb}{\end{verbatim}}

\newcommand{\suppress}[1]{}

\newcommand{\pmbeg}{\begin{pmatrix}}
\newcommand{\pmend}{\end{pmatrix}}

